# Low energy high angular resolution neutral atom detection by means of micro-shuttering techniques: the BepiColombo SERENA/ELENA sensor

S. Orsini[1], A.M. Di Lellis[2], A. Milillo[1], E. De Angelis[1], A. Mura[1], S. Selci[3], I. Dandouras[5],
P. Cerulli-Irelli[1], R. Leoni[4], V. Mangano[1], S. Massetti[1], F. Mattioli[4], R. Orfei[1],
C. Austin[5], J.-L. Medale[5], N. Vertolli[1], D. Di Giulio[1]

[1] INAF- Istituto di Fisica dello Spazio Interplanetario, Rome, Italy
[2] AMDL srl, Rome, Italy
[3] CNR - Istituto Sistemi Complessi , Rome, Italy
[4] CNR - Istituto di Fotonica e Nanotecnologie, Rome, Italy
[5] Centre d'Etude Spatiale des Rayonnements, Toulouse France

## ABSTRACT

The neutral sensor ELENA (Emitted Low-Energy Neutral Atoms) for the ESA cornerstone BepiColombo mission to Mercury (in the SERENA instrument package) is a new kind of low energetic neutral atoms instrument, mostly devoted to sputtering emission from planetary surfaces, from E ~20 eV up to E~5 keV, within 1-D (2°x76°).
ELENA is a Time-of-Flight (TOF) system, based on oscillating shutter (operated at frequencies up to a 100 kHz) and mechanical gratings: the incoming neutral particles directly impinge upon the entrance with a definite timing (START) and arrive to a STOP detector after a flight path.
After a brief dissertation on the achievable scientific objectives, this paper describes the instrument, with the new design techniques approached for the neutral particles identification and the nano-techniques used for designing and manufacturing the nano-structure shuttering core of the ELENA sensor. The expected count-rates, based on the Hermean environment features, are shortly presented and discussed.
Such design technologies could be fruitfully exported to different applications for planetary exploration.

## SCIENTIFIC AND TECHNICAL BACKGROUND

In the recent years the ENA (Energetic Neutral Atoms) detection technique has allowed new scientific investigation in the solar system. The major processes able to produce directional neutral atoms are charge-exchange and ion-sputtering.

In the first case, solar wind ions, as well as the magnetospheric plasma circulating nearby the planetary bodies, can interact with the exospheric atoms by exchanging one electron, hence producing an ENA in the energy range of several hundreds of eV up to hundreds keV. Many observations have been already done by devoted instrumentations providing charge-exchange ENA images of the interacting region in the environments of Earth (IMAGE), Mars (Mars Express), Venus (Venus Express), Jupiter and Saturn (Cassini) (e.g.: Roelof et al., 2004; Burch et al., 2003; Barabash et al., 2004, Krimigis et al., 2005). Charge-exchange ENA in the Hermean environment has been simulated (Orsini et al., 2001; Mura et al., 2005; 2006) but never observed. An ENA detector is not included in the MESSENGER payload; hence, the BepiColombo mission which includes two ENA sensors in the two spacecraft Mercury Planetary Orbiter (MPO) and Mercury Magnetospheric Orbiter (MMO) will provide a unique opportunity to obtain charge-exchange images at Mercury and to perform comparative investigations of evolution and dynamics of planetary magnetospheres.

Among the surface release processes, the ion-sputtering is particularly intriguing since the involved energies induce escape from the planet, with possible implication on its evolution. The ion-sputtering process is caused by the impact of an energetic ion (in the keV range) on the surface. The energy transfer generates a release of neutral atoms and in a minor part of ions (Hofer, 1991). The bulk of neutral emission is in the few eV range but has a significant high energy tail (Sigmund, 1969; Sieveka et al., 1984). It is a localized and highly variable release process, since the intensity of the released flux depends on the plasma precipitating flux but also on the energy of impacting ions and by the composition and the mineralogy of the target (Lammer et al., 2003).

The idea of remote-sensing in space the released neutral particles in order to map the emission from the surface was proposed for the first time for BepiColombo/MPO spacecraft to Mercury (Massetti et al., 2003; Orsini et al., 2004). Recent studies evidenced the role played by the solar wind plasma interaction with the planet in the Hermean evolution (Killen et al., 1999; Mura et al., 2005; 2006). The neutral sensor ELENA (Emitted Low-Energy Neutral Atoms), part of the SERENA instrument, was selected to acquire information about the complex surface-exosphere-magnetosphere system of the Hermean environment (Milillo et al., 2005) and to observe, in particular, the intensity, the velocity and the direction of the neutral particles escaping from planetary surface (figure 1).

In this perspective, the major ELENA scientific objectives may be resumed as in the following:

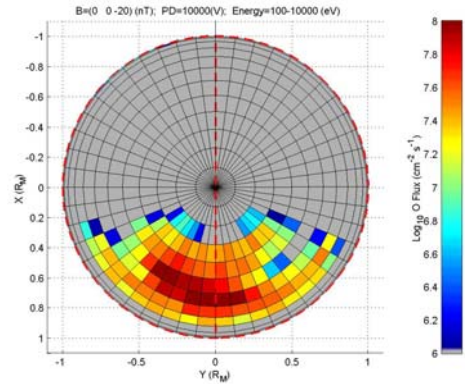

**Figure 1**: Polar map of oxygen ion sputtering escaping from the surface of Mercury. The extension of the emitting areas is strongly dependent on the actual solar wind properties (from Mura et al., 2006).

- Surface emission rate and release processes;
- Particle loss rate from Mercury's environment;
- Remote sensing of the surface composition;
- ENA imaging applications for comparative solar-planetary relationship.

The need to assess these ambitious scientific goals within a so complex environment (strongly affected by high albedo and thermal fluctuations) results in the design of an innovative ENA instrument. In fact, this unit must be extremely reliable in both angular resolution and flux sensitivity at very low energy, simultaneously able to avoid plenty of background contamination due to both photons and charged particles. Hence, the ELENA sensor of the SERENA package (Orsini et al., 2007) for the BepiColombo/MPO is innovative not only in the frame of the scientific goals achievable by ENA detection, but also for the technology necessary to detect the low energy neutral atoms sputtered from the Hermean surface (Mura et al., 2006).

From technological point of view, we have to consider that Low Energetic Neutral Atoms (L-ENA) instrumentation (few eV – hundreds eV) must be peculiar with respect to sensors dedicated to the higher energetic particles (above several hundreds eV). Actually, a major difficulty is linked to the management of the low-energetic particles, having in mind to measure energy, mass and direction for any detection (Gruntman 1997, Wurz 2000).

A commonly used method for measuring neutral atoms is to convert them into ions (by means of pass-through ultrathin foils or reflection by a conversion surface) and then to analyze the energy of the ions by means of electrostatic analyzers. The detection of secondary electrons emitted at the ionization surface/foil can be used in order to have a Start signal for Time of Flight (TOF) measurements. The STOP signal can be collected at the end of the electrostatic analyzer. This "indirect" detection technique (McComas et al., 1998) is really useful in order to have all the requested information for particles above a few hundred eV. In any case, the interaction elements (foils or surfaces) introduce uncertainty in terms of energy and angular scattering, so that the less is the particle energy, the more the information is degraded (Funsten et al., 1994).

A "direct" detection technique (without any foil / conversion surface and/or electrostatic analysis) could offer a good solution in case of fluxes at very low energy, when a fine angular resolution is required. Actually, no in-flight ENA instrument has ever been capable to successfully measure via direct detection technique, in the range of low energies.

As an application of ENA "direct" detection technique, in the following, we describe in detail ELENA, a TOF unit with a start section based on a shuttered grating system, composed by two companion membranes patterned, with slits of nanometric dimension one fixed, and the other one linked to an oscillator engine. An application of micromechanical shutter system is also ready for in-flight verification on board the Swedish Space Corporation's PRISMA satellite (PRIMA mass analyzer, Wieser et al, 2007). In the next, we will provide some details of ELENA start section with reference to the manufacturing technique. Finally, some estimates of the expected results will be reported.

## THE ELENA INSTRUMENT

### Introduction

ELENA is a Time-of-Flight (TOF) sensor, based on the state-of-the art of ultra-sonic oscillating shutters (operated at frequencies above 20 kHz and up to more than 100 kHz), mechanical gratings and Micro-Channel Plate (MCP) detectors. The purpose of the shuttering system is to digitize space and time when tagging the incoming particles without introducing "disturbing" elements, which may affect the particle's trajectory or the energy. This is particularly important in this case, where neutrals of energies of a few tens of eV must be detected.

The composite radiation made by neutrals, ions and photons impinges upon the ELENA sensor entrance (an aperture of about 1 $cm^2$), after passing through an infrared (IR) stopping grid, which reflects unwanted IR radiation to minimize the instrument heat load. The transparency for particles through the IR filter is about 50%. The ELENA sensor concept is showed in figure 2 and the instrument characteristics are summarized in table 1.

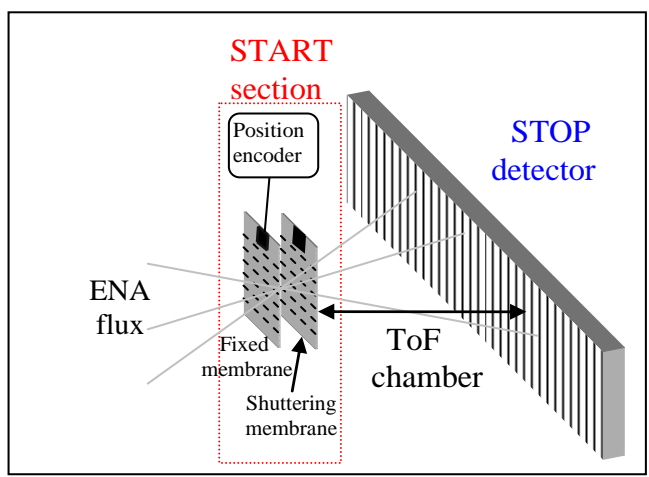

**Figure 2** ELENA sensor concept.

| ELENA characteristics | |
|---|---|
| **Energy range** | <0.02- 5 keV (mass dependent) |
| **Velocity resolution $\Delta v/v$** | Down to 15% |
| **Viewing angle** | $2°\times 76°$ |
| **Nominal angular resolution** | $2°\times 2°$ |
| **Mass resolution $M/\Delta M$** | H and heavy species |
| **Minimum integration times** | $5 \div 25$ s |
| **Geometric factor G** | ~1. $10^{-5}$ $cm^2$ sr |
| **Integral Geometric factor** | ~4. $10^{-4}$ $cm^2$ sr |

**Table 1** ELENA instrument characteristics

### The START section

The START section (see figure 2) allows the neutral particles to enter through the shuttering system with a definite timing. This section is composed principally by two elements: the shutter and the ultrasonic oscillator. This shutter has two membranes with nano-slits, one set fixed in position and the other executing a vertical shuttling motion; during the oscillating phase, the neutrals pass-through occurs only when the slits are aligned, thus determining the START time of the TOF analyzer. A capacitive control system is implemented to assure the alignment of the two membranes.

The ELENA nano-slits

The ELENA shuttering element consists of two self-standing silicon nitride ($Si_3N_4$) membranes, patterned with arrays of long and narrow openings (see figure 3a), one facing the other. These nano-slits are fabricated by electron beam lithography (EBL) and other techniques typically used for microelectronics (see figure 3b). The Institute of Photonics and Nanotechnology of CNR (Rome, Italy) is manufacturing this ELENA sub-unit. Silicon nitride has been chosen for its excellent physical proprieties (Madou, 2002): high Young's modulus, high yield strength, excellent wear resistance and low thermal expansion and conductivity. In figure 4, also the material transmittance versus strong wavelengths is shown for several film thicknesses. $Si_3N_4$ thickness of 0.5 μm is sufficient for the required opacity constraints due to the relevant environmental photon radiation.

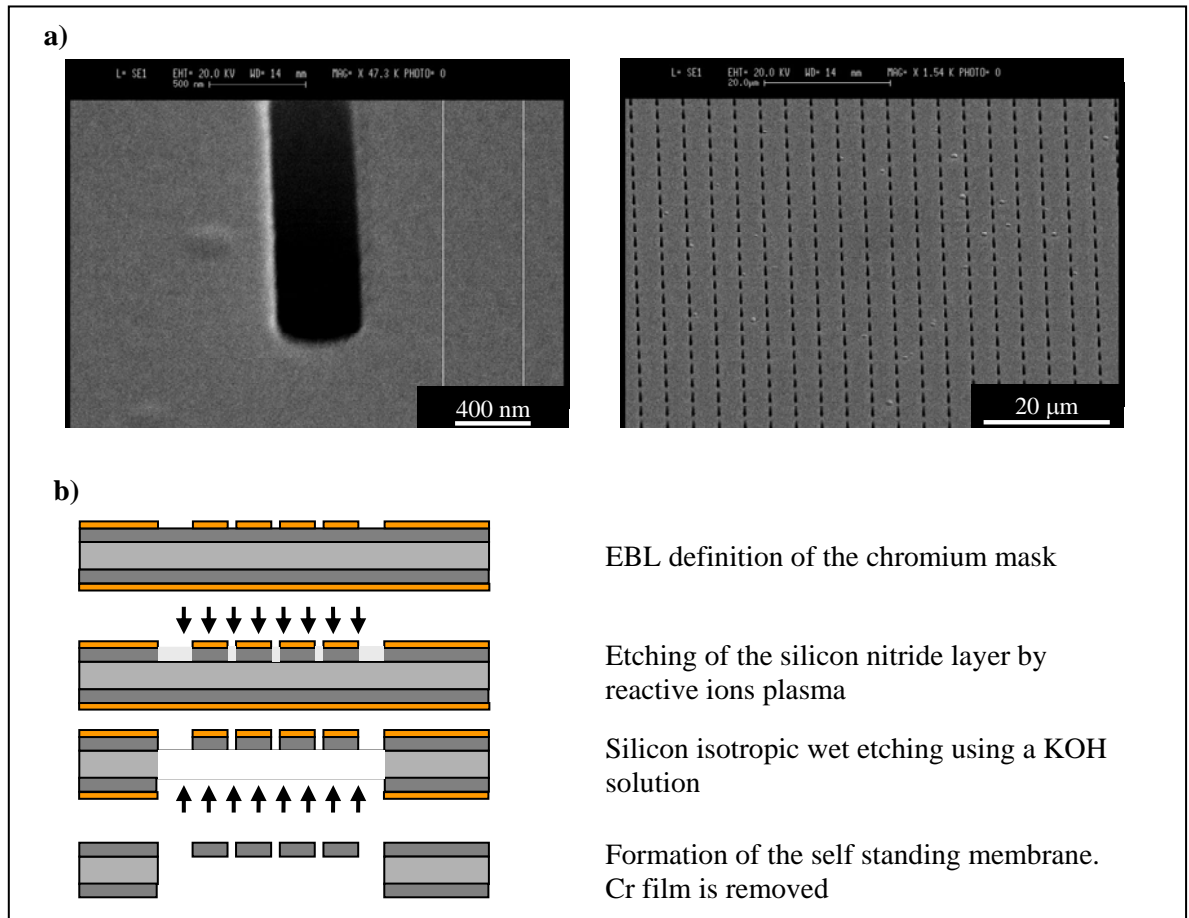

**Figure 3 a)** Scanning electron micrographs (SEM) images of one of the samples produced. **b)** Scheme of the fabrication process of the shuttering elements for the ELENA detector: from top to bottom the different layers are Cr, $Si_3N_4$, Si, $Si_3N_4$, **Cr**.

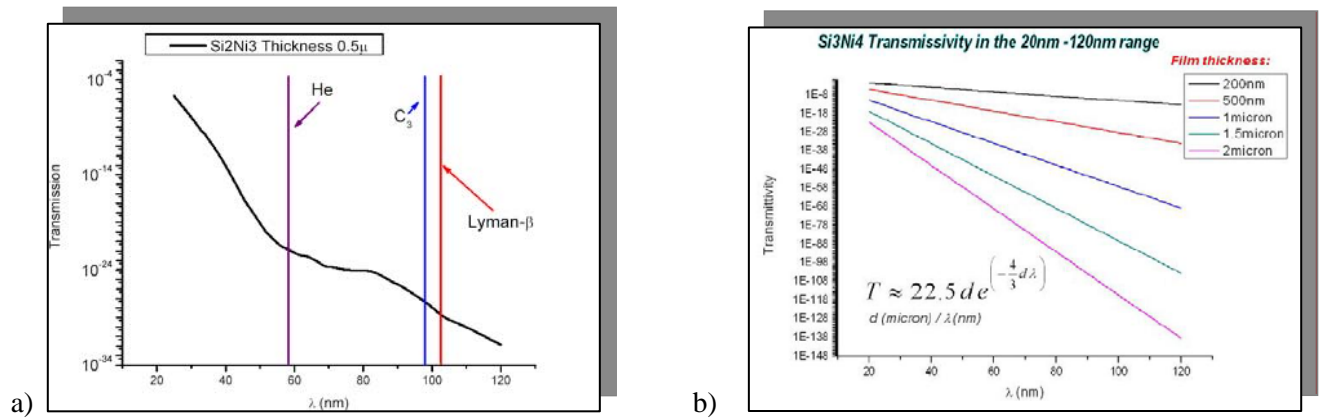

**Figure 4**: Material transmittance in the range 20 – 120 nm.
   a) $Si_3N_4$ thickness = 0.5 μm.
   b) $Si_3N_4$ thickness = 0.2 μm, 0.5 μm, 1 μm, 1.5 μm and 2 μm from top to bottom respectively.

The UV rejection is a crucial point for the ELENA instrument since photons may strongly affect the actual neutrals detection. Several nano-slits membranes have been fabricated to undergo different tests and verifications of their transmittance, performed at the Institute of Complex Systems (ISC-CNR Rome).

The transmission properties at far UV (Lyman-α region) have been investigated and encouraging results for these membranes have been obtained. Figure 5 shows the experimental rejection capability at UV wavelengths: transmittance drops out after 3 eV (about 400 nm) (Lyman-alpha ~ 121 nm, or 10 eV).

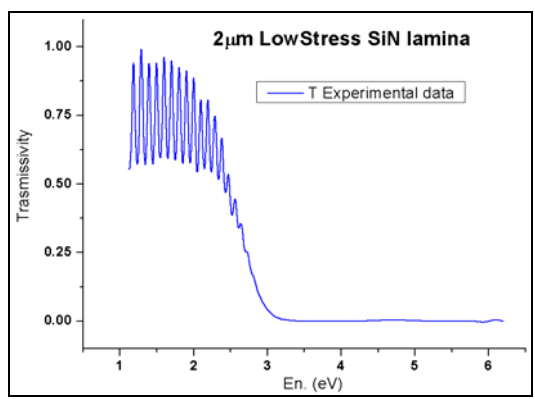

**Figure 5**: Experimental transmittance vs photon energy. Interference fringes in the Near IR side are clearly visible. Transmittance drops out after 3 eV (about 400 nm) (Lyman-alpha ~ 121 nm, or 10 eV).

The ultrasonic oscillator

Piezoelectric ultrasonic actuator has been identified as very good candidates to move the ELENA shutter, for having a compact design, fast response and low power consumption. Starting from a conventional Ultrasonic Piezo Actuator (UPA: Cedrat Technologies UPA25) a more compact system ranging up to 100 kHz can be manufactured.

Actually, for ELENA a customization of the UPA product was proposed. It is a compact, low voltage transducer for generating ultrasonic vibrations, which can be fed by a low voltage sine wave.

The position encoder

A capacitive based encoder has been studied to control the frequency and the instantaneous displacement of the ultrasonic shutter payload and the shutter phase, together with the alignment of the grids. It is based on patterned capacitive encoder, manufactured on the membranes, using an amplitude modulation-demodulation technique.

In figure 6 the equivalent electrical capacitive structure (figure 6a) is given and the concept of the fingers arrangement (figure 6b) of the capacitive encoder to be implemented on the shutter membrane for the motion control is shown.

A detailed modelling of this achievable lithography patterns has shown that capacitances in the range between 10pF and 20pF are easily achieved in the default ELENA capacitive encoder configuration. Such a varying capacitance, modulated by the actual motion of the running slits in the ELENA ultrasonic shutter, may be inserted in a "Wheatstone" bridge and then the precise mechanical phase can be retrieved an controlled by AM Amplitude Modulation techniques.

The Encoder/decoder AM demodulator has been implemented and tested in two versions, being the first based on the Analog Device AD630 part, limited in frequency for the carrier to 500kHz, and the second based on a dedicated MUX switch part operated to demodulate. The results are shown in figure 7.

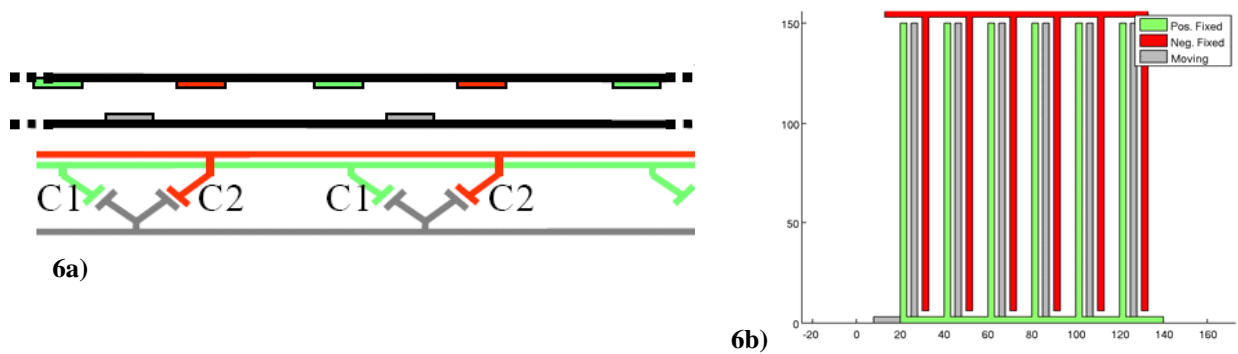

**Figure 6a-b** Capacitive encoder arrangement for shuttering control

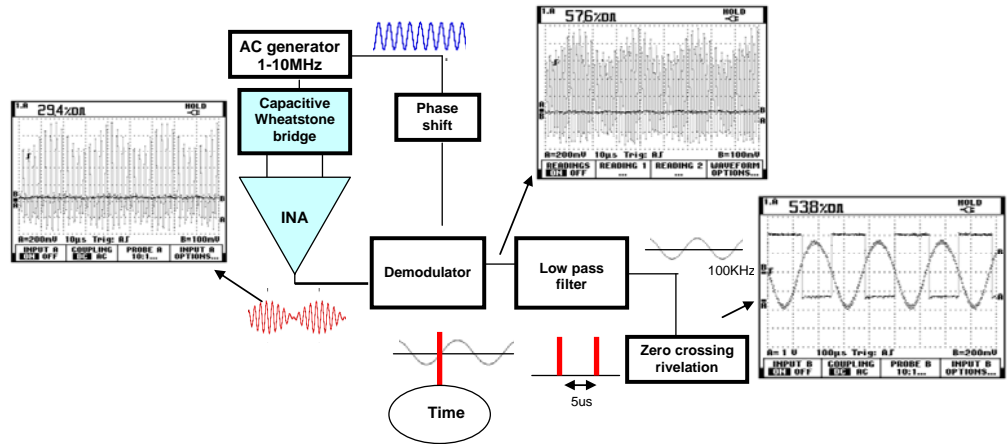

**Figure 7** Phase Encoder/decoder concept plus scope traces as measured on the implemented circuitry. Capacity variation has been emulated by modulating a reverse biased diode at bridge input.

**The TOF chamber and the STOP section**

Just after particles pass the START section, they are flown in the TOF chamber (length $L$=12 cm), where an ion deflector suppresses the charged particle flux. Finally, particles reach the STOP section (see figure 2), where they are detected by a 1-dimesional array composed by MCPs and discrete anodes set corresponding to a Field of View (FOV) of 2°x76°. The design permits an angular resolution of 2°x2° (i.e.: 38 sectors covering the whole FOV), allowing reconstructing both velocity and direction of the incoming particle.

**EXPECTED RESULTS**

**Estimated ENA flux at MPO altitude**

In the case of the BepiColombo mission, the energetic neutral particles to be detected by ELENA come primarily from ion-sputtering process, and secondarily from charge exchange (Mura et al., 2005). To estimate the neutral flux to be measured by the instrument, we assume that, during an intense solar activity, a total of $5 \cdot 10^{26}$ solar wind protons per second impact onto the surface (Leblanc et al., 2003, and references therein). These protons impact on roughly 50% of the dayside surface, and they cause the sputtering of various surface components, with a yield ($Y$) that is, on average, about 0.05 neutral particles for each incoming ion (Lammer et al., 2003), actually depending on the surface neutral specie considered. The energy distribution $f(E)$ of those sputtered neutrals peaks at few eV (Sieveka et al., 1984); nonetheless, since the energy needed to reach BepiColombo/MPO altitude is, on average, below 1 eV, it has been estimated that 90% of the emitted particles may reach the spacecraft. Moreover, it has been estimated that alphas and heavier ions can contribute to the process in a similar amount due to higher yields (Johnson et al., 1991). In summary, a maximum ENA flux $F$ of the order of $10^8$ cm$^{-2}$ s$^{-1}$ sr$^{-1}$ may be assumed at MPO periherm. Charge-exchange neutrals are emitted by the close-by exosphere, by means of interaction between solar wind protons and exospheric gas. The generated Hydrogen atoms (H-ENA) have energies of the order of 1 keV. The maximum estimated H-ENA flux is about $10^6 \sim 10^7$ cm$^{-2}$ s$^{-1}$ sr$^{-1}$, out coming from the planet limb anti-sunward where the ENA integration path is longer (Mura et al., 2005).

**ELENA detection capability**

The ENA flux will come out from the surface and from the exosphere, and the MPO satellite will be three-axes stabilized along a polar elliptical orbit (periherm=400 km; apoherm=1500 km). Hence, the ELENA FOV will point towards nadir, in order to detect the sputtering emission from the Mercury surface, with a 76° (perpendicular to the satellite orbit) by 2° instant view, and using the orbit track for exploring the signal latitudinal extension. The FOV will be tilted by 8°, so that during apoherm, the edge angular sectors will point towards the charge-exchange signal coming from the exosphere.

Background noise

Possible background arising from Sun light and cosmic rays can affect the signal. A rough estimation of the cosmic rays signal, based on flown instrumentation, produce 0.25 Counts/s (about a signal/noise = 80).

While the Ly-α (~UV band 100-150 nm) flux at 0.3 AU can be estimated from the Environment specification document of ESA as F= 0.16 W/m². Hence the photon flux is:

$$F\left[\frac{Ly\alpha}{sm^2sr}\right] = \frac{F\left[\frac{W}{m^2}\right]}{13.6 \cdot 1.6 \cdot 10^{-19} \cdot 4\pi} \frac{Ly\alpha}{sm^2sr} = 5.6 \cdot 10^{15} \frac{Ly\alpha}{sm^2sr}$$

The albedo of the planet is ~0.1 in the visible range (we apply the same overestimated factor for UV light). The Lyα signal in each channel is:

$$S_{Ly\alpha} = F \cdot 0.1 \cdot G \cdot T_{UV} \cdot \varepsilon(Ly\alpha) = 3 \cdot 10^{-4} Ly\alpha/s$$

Where $T_{UV} = 4 \cdot 10^{-8}$ is the UV transmission factor and $\varepsilon(Ly\alpha) = 2 \cdot 10^{-2}$ is the MCP efficiency to Ly-α.
Other wavelengths can produce different noise-signals, but they are estimated as negligible.

Angular resolution

One of the major merits of the sensor design is the capability for providing unprecedented performances in angular resolution within the timing discrimination constraints for the expected ENA signal. On average, most trajectories are reconstructed with accuracy below 2.5 deg.

Instrument parameters

The instrument adjustable parameters that can be tuned during the mission operation phase are the piezo oscillation frequency $\nu$ and amplitude $A$. The angular resolution ($\Delta\alpha$) of the instrument, the slit width $d$ and the TOF chamber length $L$ are parameters that also affect the performances of the instrument and can be optimized only in the design phase. The optimal configuration can be reached by maximizing the detected signal and minimizing the false counts.

The minimum detectable velocity is related to the maximum detectable time of flight $T = 1/2\nu$ since there are two apertures -START event- of the entrance for each complete shuttering cycle. Particles with longer TOF will arrive after a subsequent START event and hence they will be recorded as erroneous TOF.

The START grid is fully opened when the oscillating device is at rest (vertical *zero* position). In this configuration, each slit on the oscillating grid is aligned with the corresponding slit of the fixed grid. To ensure that during each oscillation no other opening is possible, the grid period $D$ must be at least equal to:

$$D = A + 2d \quad (1)$$

which does not take into account the finite thickness of the grids. The geometrical factor at rest position ($G_0$) relative to each angular sector is, hence:

$$G_0 = \Delta\alpha^2 \left(S\frac{d}{D}\right) f_{IR} \quad (2)$$

where $\Delta\alpha$ is approximately 2°; $S$ is the total entrance surface (1 cm²); $f_{IR}$ is the IR suppressor transparency (50%).
The theoretical opening time ($\Delta t$) can be estimated as:

$$\Delta t \approx \frac{\arcsin(d/A)}{\nu\pi} \quad (3)$$

If we multiply $G_0$ by the temporal duty cycle $r = \frac{\Delta t}{T}$, we obtain the operative $G$ of each instrument sector.

$$G = r(G_0) \quad (4)$$

The STOP efficiency ($\varepsilon$) is not considered in (4), since it depends on the energy of the particles, ranging from about $3 \cdot 10^{-3}$ at 20 eV to 0.5 at 1 keV (Stephen and Peko, 2000). The optimal configuration can be evaluated by a numerical simulation, which should take into account $\varepsilon$, the estimated flux and its energy distribution $f(E)$. At present, a good compromise can be obtained by choosing $\nu=45$ kHz, $A=1$ μm, $d\sim200$ nm, $L=12$ cm. By using these parameters, $G$ is of the order of $10^{-5}$ cm² sr, $\Delta t$ is 1.4 μs; the maximum detectable TOF corresponds to about 10 eV (for Na), $D$ is 1.4 μm. The integral $G$ over all the angular sectors is about $4 \cdot 10^{-4}$ cm² sr.

Since $\varepsilon(E)$ falls down at low energies (where the incoming flux is expected to peak), it may be necessary to integrate the data on *n* contiguous angular sectors, and during a longer time $T_{int}$. Anyway, since the real condition at Mercury is not known, ELENA will maintain the possibility to have a spatial resolution of $2°\times 2°$, allowing an imaging of the surface with resolution ranging between 15 and 55 km for peri- and apo-herm, respectively, and $T_{int}$ between 5 s and 25 s depending on the spacecraft velocity along the orbit. The edge sectors, which will observe the limb when the MPO spacecraft will approach the apoherm, will provide observations of the ENA of SW origin.

Ion–sputtering signal simulations

Figure 8 shows a simulation of the energy-integrated (between 20-1000 eV) sputtered O from vantage point close to the periherm, on the day-side. The instantaneous FOV of the linear array of the ELENA sensor is a single row of the image. The spacecraft footprint track will provide the second dimension. The overall picture results from time integration along the MPO orbit, which allows latitude signal tracking.

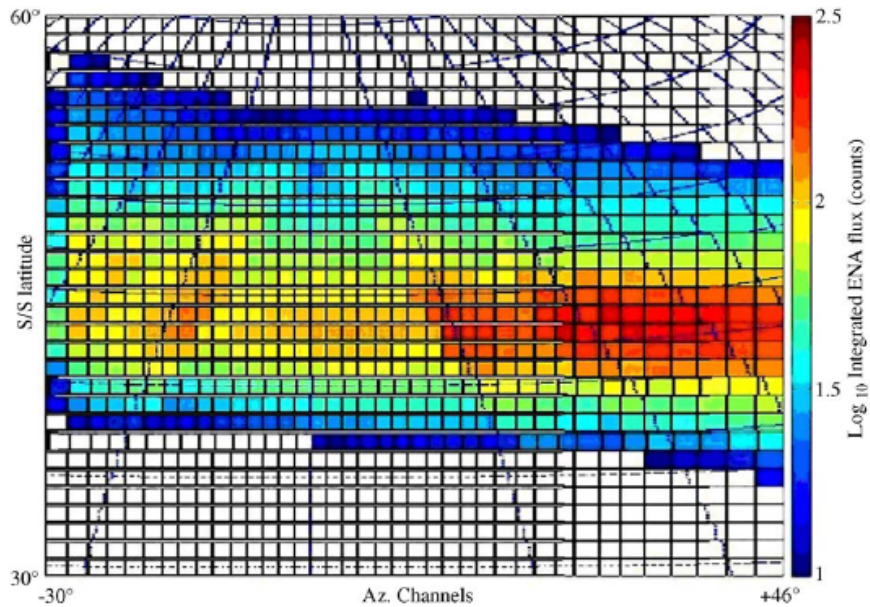

**Figure 8**. Energy-integrated (between 20-1000 eV) sputtered O from vantage point close to the periherm, on the day-side (from Mura et al., 2006)

Figure 9 shows both an analytical and numerical estimation of the count rates related to neutral fluxes (where we consider that all detectable particles are Na) that can be measured by ELENA instrument on the planetary dayside periherm. We have defined a number of TOF channels, and we have accumulated the simulated events in these TOF channels with a MonteCarlo model, assuming a sputtering distribution of a number of particles equal to $T_{int}\, G\, F\, n$. $T_{int}$ and *n* has been chosen so that the projection of the MPO track during $T_{int}$ is equal to the surface resolution (in order to have a square pixel on the surface). Hence, $T_{int} = 90$ s and $n=5$; in this case, the angular resolution of the analysis is $n\Delta\alpha=10°$. We have applied the proper STOP detection efficiency $\varepsilon(E)$ to the simulated signal for each TOF event.

The left panel shows the analytical prediction of the measurable signal (red columns) and particles arriving after the new opening (grey columns). The right panel shows the MonteCarlo simulation (green columns; blue columns represent the statistical error).

The good agreement between the two panels of Figure 9 demonstrates that the count rates are high enough to allow reconstruction of the incoming fluxes.

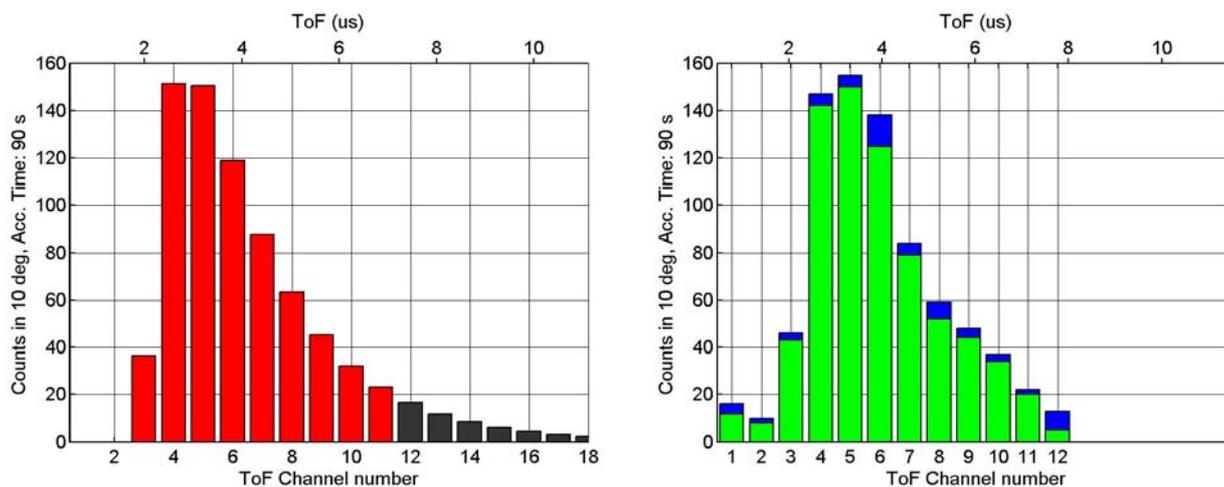

**Figure 9**. *Left panel*: analytical prediction of the measurable signal (red columns) and particles arriving after the new opening (gray columns). Gray channels are "virtual", since their counts are actually collected in channels 1-5.
*Right panel*: MonteCarlo simulation of the TOF signal (blue and green columns). (from Mura et al. 2007)

Charge-exchange signal simulations

Charge-exchange H-ENA have typical *TOF* of the order of $T_H$=200 ns. Since this is much shorter than $\Delta t$ (~1.4 μs), set in the *ion-sputtering mode*, it is likely that most of the H-ENAs (90%) arrives to the STOP surface while the START is still open and the STOP surface is still illuminated by UVs. Such particles will not be detected by ELENA. However, it is possible to increase the shuttering frequency $\nu$ to partially take into account this fact. If the highest frequency reachable by the shuttering is of the order of $\nu$ =100 kHz, 25% of the particles entered will be measured. By taking into account *G* and the estimated H-ENA flux, up to about 50 particles per second can be measured by the angular sectors of ELENA looking toward the ENA generating region, mainly the planet limb.

In s/c positions and altitudes, where both signal are foreseen, we will set the ELENA frequency in order to detect sputtered neutrals, since surface emissivity is our primary scientific objective. Anyway, when the spacecraft will be in the night side at apoherm it is believable that only charge-exchange particle will arrive at the detector, thus permitting to set the highest frequency. In figure 10 the simulation of H- and O- ENA fluxes with similar energy distributions (~1 keV) detectable by ELENA shows that the possibility to discriminate these two major species.

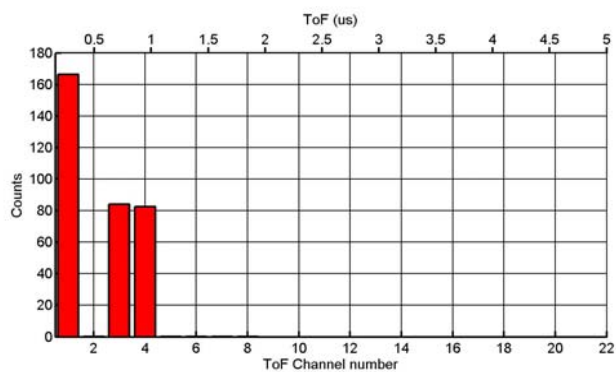

**Figure 10** ELENA mass resolution capability based on TOF analysis. Simulation of the H- and O- ENA fluxes with similar energy distributions (~1 keV) detectable by ELENA The shutter frequency has been set to 100 kHz (from Mura et al. 2007).

**CONCLUSION**

The measurement of Low Energetic Neutral Atoms has a lot of technological implications. The ELENA instrument with the TOF system based on the oscillating shutter answers to some of the difficulties to measure particles in the energy range below 1 keV. It is a new application of the straightforward TOF system with chopper element with the necessary devices and parameters study for space application. This technique is particularly suitable to measure sputtered particles from surface of Mercury. Anyway, these design technologies can be used in different planetary explorations and low energy range particles investigation.

Acknowledgments. The authors thank the two referees for their valuable contribution and help in making the paper more reliable and complete. SERENA/ELENA is financially supported by the Italian Space Agency (ASI) under contract number I/090/06/0, with some contribution from CNES (F).